\documentclass[epj]{svjour}
\usepackage{graphics,amssymb,color,bm}
\newcommand{\gp}{\dot{\gamma}}
\newcommand{\is}{s$^{-1}$}

\begin{document}
%
\title{Surfactant micelles: model systems for flow instabilities of complex fluids}
\author{Christophe Perge \and Marc-Antoine Fardin \and S\'ebastien Manneville
%
}                     
\offprints{sebastien.manneville@ens-lyon.fr}          
\institute{Laboratoire de Physique, \'Ecole Normale Sup\'erieure de Lyon, 46 all\'ee d'Italie, 69364 Lyon cedex 07, FRANCE}
\date{Received 15 July 2013 / Revised version: date}
%
\abstract{Complex fluids such as emulsions, colloidal gels, polymer or surfactant solutions are all characterized by the existence of a ``microstructure'' which may couple to an external flow on timescales that are easily probed in experiments. Such a coupling between flow and microstructure usually leads to instabilities under relatively weak shear flows that correspond to vanishingly small Reynolds numbers. Wormlike micellar surfactant solutions appear as model systems to study two examples of such instabilities, namely shear banding and elastic instabilities. Focusing on a semidilute sample we show that two dimensional ultrafast ultrasonic imaging allows for a thorough investigation of unstable shear-banded micellar flows. In steady state, radial and azimuthal velocity components are recovered and unveil the original structure of the vortical flow within an elastically unstable high shear rate band. Furthermore thanks to an unprecedented frame rate of up to 20,000 fps, transients and fast dynamics can be resolved, which paves the way for a better understanding of elastic turbulence.
} 
\maketitle
%

\section{Introduction: flow instabilities without inertia}
\label{s.intro}
A complex fluid is characterized by its {\it microstructure}, i.e. a supramolecular organization involving one or several length scales intermediate between the molecular scale and the container size  \cite{Larson:1999}. For instance in an oil-in-water emulsion, this {\it mesoscopic} length scale is simply given by the oil droplet diameter, usually 1--10~$\mu$m. Depending on the material the mesoscopic scale may range from tens of nanometers (diameter of colloidal particles or radius of gyration of polymer molecules) up to several millimeters (granular pastes or bubble diameter in a foam). It is generally measured through scattering techniques such as small angle light, neutron, or X-ray scattering \cite{Larson:1999}.

Due to its microstructure a complex fluid shows significant {\it viscoelasticity}: when weakly perturbed from equilibrium, its mechanical response involves both a solid-like behaviour that results from the elastic recoil of the mesoscopic elements and a liquid-like component that results from the fact that these elements may rearrange plastically and ultimately flow. Such viscoelastic properties are generally measured quite easily, especially since the microstructure relaxation times after a small deformation are macroscopic and typically fall into the range $\tau\simeq 10^{-2}$--$10^2$~s. The orders of magnitude of the elastic modulus at rest and of the zero-shear viscosity are respectively $G_0\simeq 1$--$10^4$~Pa and $\eta_0\simeq 1$--$10^4$~Pa.s. When the material microstructure is constituted of an amorphous, compact assembly of jammed elements, relaxation times may become much longer than the actual duration of the experiment used to measure viscoelastic properties, typically larger than $10^2$~s. Such materials were coined ``soft glassy materials'' \cite{Weeks:2007}.

\begin{figure*}
\begin{center}
\resizebox{0.9\textwidth}{!}{\includegraphics{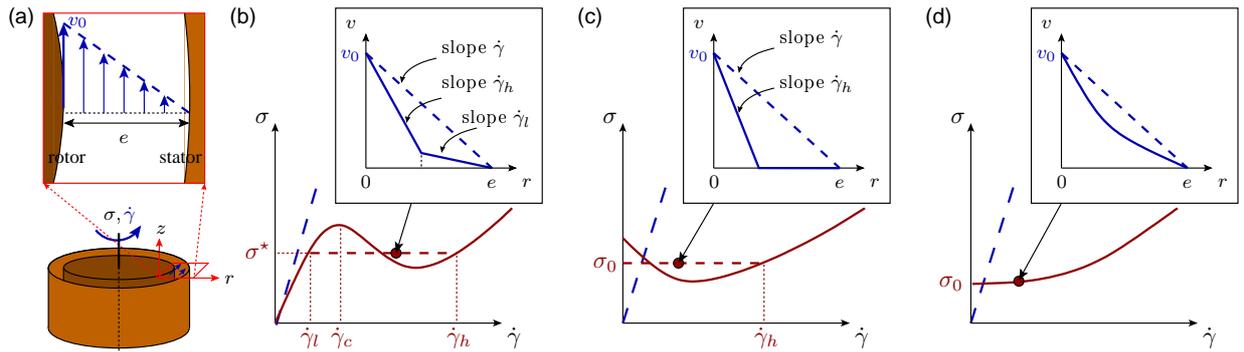}}
\caption{(a) Shear flow in the Taylor-Couette geometry and notations as used in the text. Flow curves $\sigma$ vs $\gp$ and velocity profiles $v(r)$ expected in (b) a semidilute wormlike micelle solution, (c) a yield stress material with shear localization, and (d) a ``simple'' yield stress material without shear localization. Blue dotted lines indicate the case of a Newtonian fluid.}
\end{center}
\label{f.rheol}
\end{figure*}

Besides viscoelasticity close to mechanical equilibrium, known as {\it linear} viscoelasticity, complex fluids show a wide variety of fascinating properties when submitted to large deformations and/or steady flow. For instance, under a constant shear rate, denoted as $\gp$ in the following [see Fig.~\ref{f.rheol}(a) for a sketch of the Taylor-Couette (TC) geometry used in our experiments], the fluid microstructure may reorganize, leading to a strong dependence of the viscosity $\eta$ on the applied shear rate \cite{Larson:1999}. In general the viscosity decreases with $\gp$ (shear-thinning behaviour) as is the case for polymer solutions where the shear flow stretches and aligns the molecules thus lowering viscous friction. Many other {\it non-Newtonian} behaviours are commonly observed that result from the coupling between microstructure and deformation. Shear-thickening is the phenomenon by which flow may induce a viscosity increase  e.g. through shear-induced cluster formation in dense colloidal suspensions \cite{Foss:2000,Maranzano:2002} or through particle migration and concentration gradients in granular pastes \cite{Fall:2010}. The yield stress behaviour, which is ubiquitous in soft glassy materials, is yet another manifestation of the underlying microstructure. The first goal of {\it nonlinear rheology} is therefore to probe the flow--microstructure coupling through the determination of the flow curve, i.e. the shear stress $\sigma=\eta\gp$ vs shear rate $\gp$ curve \cite{Macosko:1994}. Figures~\ref{f.rheol}(b)--(d) show classical examples encountered in the systems discussed below.

In a typical nonlinear rheological experiment performed at $\gp=10$~\is\ over a sample of thickness $e=1$~mm and of viscosity $\eta\simeq 1$~Pa.s, the Reynolds number is $Re\equiv \rho \gp e^2/\eta\simeq 10^{-2}$, where $\rho\simeq 10^3$~kg.m$^{-3}$ is the fluid density. For such a low Reynolds number inertia is negligible and hydrodynamic instabilities, such as the Taylor-Couette instability, classically observed in simple fluids sheared in the TC geometry, do not develop in complex fluids. Still, many ``instabilities at zero Reynolds number'' have been discovered in complex fluids over the last 60 years~\cite{Larson:1992,Morozov:2007}. Most of these new phenomena were first reported in polymer solutions, where {\it normal forces} develop perpendicular to the velocity direction and play a major role. For instance it has long been known that a polymer solution may climb along a rotating rod up to impressive heights. This ``rod-climbing'' effect was given a detailed explanation by Weissenberg as early as 1946 \cite{Weissenberg:1946}. Polymers extruded under moderate pressure drops within a narrow pipe are commonly seen to swell at the pipe exit contrary to simple fluids \cite{Tanner:1970}. In some strongly viscoelastic polymer solutions sheared in the TC geometry, purely elastic instabilities are reported with a phenomenology that is strikingly similar to the classical inertial Taylor-Couette instability: transition from a laminar flow to a toroidal secondary flow made of counter-rotating vortices stacked along the vorticity direction, secondary instabilities, and subsequent transition to ``elastic turbulence'' \cite{Larson:1990,Groisman:2004}. In this case, the relevant dimensionless number has been consistently shown to be the Weissenberg number ~\cite{Morozov:2007}, defined as $Wi \equiv \tau\gp$, where $\tau$ is the relaxation time of the microstructure rather than the viscous dissipation time $\tau_i=\rho e^2/\eta$ (that would give the Reynolds number). In both the purely inertial and purely viscoelastic cases, for a small gap TC geometry, the precise instability threshold depends on a Taylor number, either inertial $Ta_i=\sqrt{e/R_i}\, Re$ or elastic $Ta=\sqrt{e/R_i}\, Wi$, where $e$ is the thickness of the unstable sheared fluid and $R_i$ is the inner radius of the TC geometry.  

Moreover as in simple fluids, the presence of boundaries and confining walls may lead to flow instabilities in complex fluids. However, in this case again, instabilities do not arise from inertia but from elasticity or from wall slip through the development of lubricating layers. The ``sharkskin instability'' in extrusion \cite{Denn:2001} or the ``spurt effect'' in pipe flow \cite{Vinogradov:1973} are two examples of such wall-driven instabilities in complex fluids. Finally as we shall see in the next section, the coupling between the microstructure and the flow can itself be a source of instability without inertia. 

The goal of this paper is to show that the issue of flow instabilities in complex fluids can be addressed by focusing on the case of a single model system, namely surfactant wormlike micelles. In the next section, we recall the structure and rheology of wormlike micellar solutions. By briefly reviewing the current literature, we emphasize that such systems could display the full range of instabilities described above when the surfactant concentration and/or temperature is varied. In Section~\ref{s.usv} we study in more detail a semidilute sample by means of ultrafast ultrasonic imaging coupled to standard rheology. Velocity maps are presented with a high frame rate, such that transient regimes and fast unstable dynamics can be resolved. These new data provide quantitative confirmation that in this particular system Taylor-like vortices are present in the high shear rate band as soon as shear banding occurs. The structure of these vortices can be accessed as well as their long-term dynamics which involve annihilation and creation processes. In the turbulent regime, two-dimensional imaging appears as a promising tool for the space-time characterization of elastic turbulence.

\section{The structure and rheology of wormlike micelles}
\label{s.rheol}

Depending on temperature, concentration, and salt content, surfactant molecules in aqueous solution can spontaneously arrange into long, cylindrical, semiflexible micellar aggregates that can reach several micrometers in length for a diameter of about 10~nm. Such self-assembled aggregates are known in the literature as {\it wormlike micelles} and a number of reviews are available on these surfactant systems \cite{Berret:2005,Cates:2006,Lerouge:2010}. These materials are widely used in the industry for oil recovery as well as in detergents \cite{Yang:2002}. Contrary to polymer molecules whose length is fixed by the chemical polymerization reaction, wormlike micelles have the ability to break and recombine constantly under thermal agitation. This peculiar behaviour yields model rheological properties that depending  on the concentration regime involve most of the ``zero Reynolds number'' instabilities listed above. In the following we distinguish between dilute, semidilute, and concentrated regimes and review a few classical shear-induced effects for each of these regimes.

\subsection{Dilute regimes: shear-induced structures}
At low concentrations, from a few hundred ppm to about 0.1-1~wt.$\%$ (yet far above the critical micellar concentration), wormlike micelles are short, do not interact and essentially display Newtonian behaviour at low shear rates with a viscosity of a few mPa.s~\cite{Lerouge:2010}. However above some critical shear rate, significant shear-thickening is often reported in dilute systems \cite{Rehage:1982,Hu:1998a}. Such a surprising behaviour has been attributed to shear-induced growth and fusion of wormlike micelles that lead to a gel-like shear-induced structure (SIS) \cite{Boltenhagen:1997a,Berret:2002}. For instance, Pine and coworkers identified a shear-induced gel in a TTAA and NaSal system sheared in a TC device \cite{Hu:1998a,Boltenhagen:1997b}. Due to the highly viscoelastic nature of the shear-induced gel, the flow in the shear-thickening regime usually involves wall slip and fractures \cite{Hu:1998b}. At steady state, the stress displays large fluctuations. At high shear rates the gel is destroyed and low viscosities are recovered. 

The nature of the flow states remains elusive in dilute systems. This is partly due to the fact that reproducibility is usually poor at low concentrations where impurities may perturb the experiments. But maybe more importantly, the analysis is intrinsically limited by the fact that (i)~both inertial and viscoelastic instabilities can be at play, since the viscosity are sufficiently low to lead to large values of $Re$, and (ii)~neither the microstructure relaxation time scale $\tau$ of the shear-thickened state nor its viscous dissipation time $\tau_i$ can be given by linear rheology measurements. 

The first problem is shared by dilute polymer solutions. For simple fluids, without viscoelasticity, the relevant dimensionless number can only be $Re$, since $Wi=0$. For complex fluids in the limit of purely viscoelastic instability, the dimensionless number is $Wi$, since $Re\simeq 0$. Yet, in general both $Re$ and $Wi$ can be relevant~\cite{Muller:2008}. The addition of small concentrations of polymer, or weak viscoelasticity, stabilizes Taylor-Couette flow against the formation of inertial Taylor vortices. That is, the critical Reynolds number in a given geometry increases relative to a Newtonian fluid. This flow stabilization for weak elasticity is also used in turbulent drag reduction applications, with polymers, but also with dilute micellar solutions~\cite{Shenoy:1984}. Nevertheless, in the limit $Wi \gg Re$, the flow becomes unstable again via mechanisms akin to the purely viscoelastic case. In between these two limits, the crossover between instabilities dominated by inertia and those dominated by elasticity remains to be fully understood. For recent progress on the inertio-elastic crossover in TC flow, one can read Dutcher and Muller's latest study and references therein~\cite{Dutcher:2013}. For a recent focus on turbulent drag reduction, the reader is referred to Samanta \textit{et al.}~\cite{Samanta:2013}.

The second problem is usually avoided in recent studies on polymer solutions, where the polymers are chosen such that the time scales $\tau$ and $\tau_i$ do not depend on the shear rate $\gp$, in other words, by considering Boger fluids~\cite{Muller:2008}. Nonetheless, earlier studies had unclear rheological characterization and were thus potentially affected by the same problem~\cite{Larson:1992,Muller:2008}. For dilute micellar solutions, the emergence of a SIS prevents the design of any ``micellar Boger fluid''. Shear-thickening modifies $\eta$ and thus $\tau_i$. Besides the underlying SIS has a relaxation time $\tau$ that strongly depends on the applied shear rate $\gp$. Since the SIS disappears after flow cessation, regular linear rheology methods cannot be used to estimate $\tau$. Large amplitude oscillatory shear~\cite{Hyun:2011} or superposition rheology~\cite{Dhont:2001,Ballesta:2007,Kim:2013} could be used instead. Recent studies on the flow of dilute micellar solutions in porous media managed to generate irreversible flow induced structured phases~\cite{Vasudevan:2010,Cardiel:2013}. Such phases can then be probed by small angle oscillatory tests, but it may have different properties than the usual reversible SIS.  

A few recent studies also investigated the impact of the rheology of dilute micellar solutions on interfacial (rather than bulk) instability mechanisms, such as the Faraday instability \cite{Ballesta:2006}, or the Plateau-Rayleigh instability~\cite{Boulogne:2013}. Still much more research effort is required to fully understand instabilities in dilute wormlike micelles. We believe that space- and time-resolved techniques similar to that used in the present article for semi-dilute solutions should be extended to tackle the dilute case. Outer rotation in TC geometry could also be used to differentiate between inertial and viscoelastic instability mechanisms, since only the inertial instability is bound to the Rayleigh criterion (see \cite{Dutcher:2013} for details). 

At the extremity of the dilute concentration range, around concentrations of about 1~wt.$\%$, there exists a small crossover regime with semi-dilute shear-banding solutions described below. This regime can exhibit a mixture of the dynamics described here and in the next section. The understanding of this crossover is currently limited by our understanding of the dilute and semi-dilute regimes. For instance the equimolar solution made of cetylpyridinium chloride and sodium salicylate, the concentration of each component being fixed at 40 mM, pertains to the crossover regime. This particular system has been extensively studied, especially by Fischer and coworkers \cite{Herle:2005,Herle:2007,Herle:2008}. For the original study, see Wheeler \textit{et al.}~\cite{Wheeler:1998} and for the most recent study, see Lutz \textit{et al.}~\cite{Lutz:2013}.

\subsection{Semidilute and concentrated regimes: shear-banding and elastic instabilities}

\begin{figure}
\begin{center}
\resizebox{\columnwidth}{!}{\includegraphics{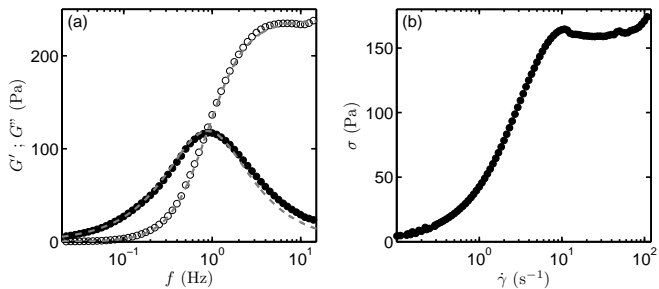}}
\caption{\label{f.rheology}Standard rheological features of a wormlike micelle solution ([CTAB]=0.3 M, [NaNO$_3$]=0.4 M) seeded with hollow glass spheres at 1~wt.$\%$. (a) Linear viscoelastic moduli $G'$ (open symbols) and $G''$ (filled symbols) as a function of oscillation frequency $f$ for a deformation amplitude of $1\%$. Gray dashed lines correspond to a Maxwellian behaviour with $G_0=238$~Pa and $\tau=0.18$~s. (b)~Flow curve $\sigma$ vs $\gp$ obtained by sweeping up the shear rate with logarithmic sampling (5~s per point and averaged over the last 2~s). Measurements performed in a cone-and-plate geometry (diameter 40~mm and angle 2$^\circ$) at $T=30^\circ$C.}
\end{center}
\end{figure}

In the semidilute and concentrated regimes, wormlike micelles entangle into a strongly viscoelastic network. These regimes typically occur for concentrations ranging from 0.1-1~wt.$\%$ to $10-20$~wt.$\%$. The peculiar breaking and recombination dynamics of micelles result in an almost perfect Maxwellian behaviour for small deformations, characterized by a single relaxation time $\tau\simeq 0.1$--1~s and a high-frequency elastic modulus $G_0\simeq 10$--100~Pa (and hence a zero-shear viscosity $\eta_0=G_0\tau\simeq 1$--100~Pa.s)~\cite{Berret:2005,Cates:2006,Lerouge:2010}. Viscoelastic measurements in the linear regime performed on a typical semidilute wormlike micellar system are shown in Fig.~\ref{f.rheology}(a).

Turning to nonlinear rheology, semidilute wormlike micelles show a weakly shear-thinning behaviour at low shear rates: like classical polymers, they progressively stretch and align in the flow as $\gp$ is increased. However, above a critical shear rate $\gp_c$, the micelles suddenly disentangle and align, which leads to a nematic-like state. This abrupt transition can be seen as a {\it mechanical instability} or an {\it out-of-equilibrium phase transition}~\cite{Fardin:2012a}. It is characterized by a portion of the flow curve $\sigma$ vs $\gp$ that displays a negative slope as depicted in Fig.~\ref{f.rheol}(b). Such an instability, which occurs at virtually vanishing Reynolds number, was predicted theoretically in the early 1990s \cite{Spenley:1993}. Experimentally the unstable zone shows as a characteristic {\it stress plateau} at a well-defined shear stress $\sigma^\star$. This stress plateau, where the viscosity thus decreases as $\eta\sim\gp^{-1}$, separates two increasing branches for $\gp<\gp_l$ and $\gp>\gp_h$ respectively \cite{Berret:2005,Cates:2006,Lerouge:2010,Fardin:2012a}. Figure~\ref{f.rheology}(b) presents an example of such a flow curve showing a stress plateau.

A simple analogy with a first-order equilibrium phase transition --where the low (high resp.) shear rate branch corresponds to the state of entangled (aligned resp.) mi\-celles-- allows one to expect ``phase coexistence'' for $\gp_l<\gp<\gp_h$ in the form of a spatially heterogeneous flow constituted of two (or more) regions bearing respectively the local shear rates $\gp_l$ and $\gp_h$. Such a heterogeneous flow, termed ``shear banded'' flow, is sketched in Fig.~\ref{f.rheol}(b). Although this simple description of the {\it shear-banding instability} obviously fails in providing any selection criterion for the stress plateau at $\sigma^\star$, it turns out to be quite efficient and triggered lots of theoretical works that fall out of the scope of the present discussion \cite{Cates:2006,Fardin:2012a,Olmsted:2008}. In particular, on the experimental side, it allows to interpret some transient rheological measurements in terms of metastability \cite{Grand:1997,Berret:1997b}. Depending on the sweep rate used for the flow curve measurements, the stress plateau may be preceded by a stress maximum reminiscent of metastable states as illustrated in Fig.~\ref{f.rheology}(b) for $\gp\simeq 10$~\is.

Thanks to the development of various techniques enabling a local investigation of the velocity field during rheological measurements \cite{Manneville:2008}, a shear-banded flow along the stress plateau was first clearly evidenced in the cone-and-plate geometry \cite{Britton:1997} and later quantitatively shown to agree with theoretical predictions in the TC geometry \cite{Salmon:2003c}. Therefore the above simple scenario for shear banding is now experimentally and theoretically well-established in a large variety of wormlike micelle solutions. Shear banding as the result of a shear-induced transition has also been found in a number of other complex fluids ranging from lamellar phases \cite{Salmon:2003d} to copolymers \cite{Manneville:2007} and associative polymers \cite{Sprakel:2008}. However, recent years have seen growing evidence that the simple scenario at best holds only when time-averaged flow is considered in the steady-state. Indeed time-resolved approaches have shown that shear banding most often goes along with strong velocity fluctuations not only in wormlike micelles \cite{Lerouge:2010,Becu:2004,Becu:2007,Lopez:2004} but also in other systems \cite{Manneville:2007,Sprakel:2008,Manneville:2004b}. Using rheo-optical measurements Lerouge and coworkers interpreted these fluctuations as the consequence of a purely elastic instability in the aligned state of the micelles (high-shear band) \cite{Fardin:2009,Fardin:2011,Fardin:2012c}. While the centrifugal force triggers the inertial Taylor-Couette instability in a simple fluid, the elastic instability is generated by centripetal normal forces that arise due to the curvature of the streamlines in the TC geometry \cite{Pakdel:1996}. In the case of a shear-banded flow the instability only develops in the high-shear region. This results in a specific unstable flow where Taylor-like vortices deform the interface that separates the two shear bands \cite{Fardin:2012d}. The aim of Sect.~\ref{s.usv} below is to provide new quantitative evidence for such a scenario based on velocity measurements in the TC geometry though ultrafast ultrasonic imaging.

\subsection{Highly concentrated regimes: wormlike micelles with a yield stress}

When the surfactant concentration is increased deep into the entangled regime wormlike micelles start to strongly interact through steric interactions leading to a succession of nematic and other ordered  phases typical of liquid crystals (hexagonal, cubic, lamellar, etc.)~\cite{Lerouge:2010}. The equilibrium isotropic/nematic (I/N) transition occurs for concentrations around $20-30$~wt.$\%$, where\-as the transition to, say, a hexagonal phase occurs around $30$~wt.$\%$. In general, both transitions are of the first order and biphasic regions exist in the equilibrium phase diagram depending on surfactant and salt concentrations, and temperature. 

The I/N transition line has been extensively studied in the 1990s by Berret \textit{et al.}~\cite{Berret:2005}. If a surfactant system is isotropic but close to its equilibrium I/N transition at rest, then shear leads to shear-banding with a high shear rate band with a nematic order~\cite{Berret:2005,Lerouge:2010}. In contrast, in the semi-dilute regime, far from the transition line, the microstructure of the high shear rate band may be more complex~\cite{Lerouge:2010}. Nonetheless, more recent studies focusing on the main azimuthal flow field rather than on the microstructure have not revealed any major differences in the macroscopic dynamics of shear-banding flows for systems below or above the I/N transition~\cite{Helgeson:2009a,Helgeson:2009b}. In the most recent study, we also did not report any major difference in the spatiotemporal dynamics of the secondary vortex flow~\cite{Fardin:2012c}. Nevertheless, well into the nematic regime, it is known that shear-banding disappears and is replaced by flow-aligning and tumbling instabilities~\cite{Lerouge:2010}. The fate of viscoelastic instabilities in this regime is completely unexplored. A careful assessment of the effect of the I/N equilibrium transition on viscoelastic instabilities of micellar flows is one of the future applications of the ultrasonic technique used here for a semidilute system.           

The more ordered regimes (hexagonal, cubic, etc.) remain unfortunately understudied. In these regimes, wormlike micelles can present a solidlike behaviour at rest and can develop a yield stress as described in the introduction \cite{Larson:1999,Lerouge:2010,Schmidt:1998,Montalvo:1996}. In this case the flow curve of concentrated wormlike micelles can be seen as the limit of that of semidilute systems when $\gp_l\rightarrow 0$ as sketched in Fig.~\ref{f.rheol}(c). The simple analogy with a first-order phase transition described above thus predicts the coexistence of a flowing phase sheared at $\gp_h$ and an unsheared solidlike region ($\gp_l=0$) when a shear rate is applied below $\gp_h$. Above $\gp_h$ a homogeneous flow is expected as the whole fluid has become fluidlike. This type of shear banding instability, sometimes referred to as {\it shear localization} in the literature of soft jammed materials, has been reported in colloidal gels \cite{Moller:2008} or clay suspensions \cite{Coussot:2002a} and often involves strong wall slip effects and thixotropy.

When the limit $\gp_h\rightarrow 0$ is further considered, one is left with a flow curve that monotonically increases from the yield stress $\sigma_0$ [see Fig.~\ref{f.rheol}(d)]. This corresponds to the ``simple yield stress'' behaviour classically modeled by a Bingham ($\sigma=\sigma_0+\eta\gp$) or Herschel-Bulkley law ($\sigma=\sigma_0+A\gp^n$) \cite{Ragouilliaux:2007,Moller:2009b,Coussot:2010}. In this case a homogeneous flow is predicted whatever the imposed shear rate. Note that the velocity profile in TC geometry can be significantly curved as sketched in Fig.~\ref{f.rheol}(d) due to the very strong apparent shear-thinning close to the yield stress and to the stress heterogeneity inherent to the TC device. In large-gap TC geometries, the local shear stress can even become smaller than the yield stress next to the outer wall. This leads to apparent shear localization which is only due to the large curvature of the TC device and should not be confused with that observed in the case of an underlying shear-banding instability as in Fig.~\ref{f.rheol}(c) \cite{Moller:2009b,Schall:2010}.

As far as surfactant systems are concerned, the particular yield stress fluid obtained (either simple or thi\-xo\-tro\-pic) and the possibility of viscoelastic flow instability remain to be explored. Here again, the two-dimensional flow imaging technique used below could be instrumental. 

\section{Ultrasonic imaging of shear banding and elastic instability in semidilute wormlike micelles}
\label{s.usv}
\begin{figure*}
\begin{center}
\resizebox{0.7\textwidth}{!}{\includegraphics{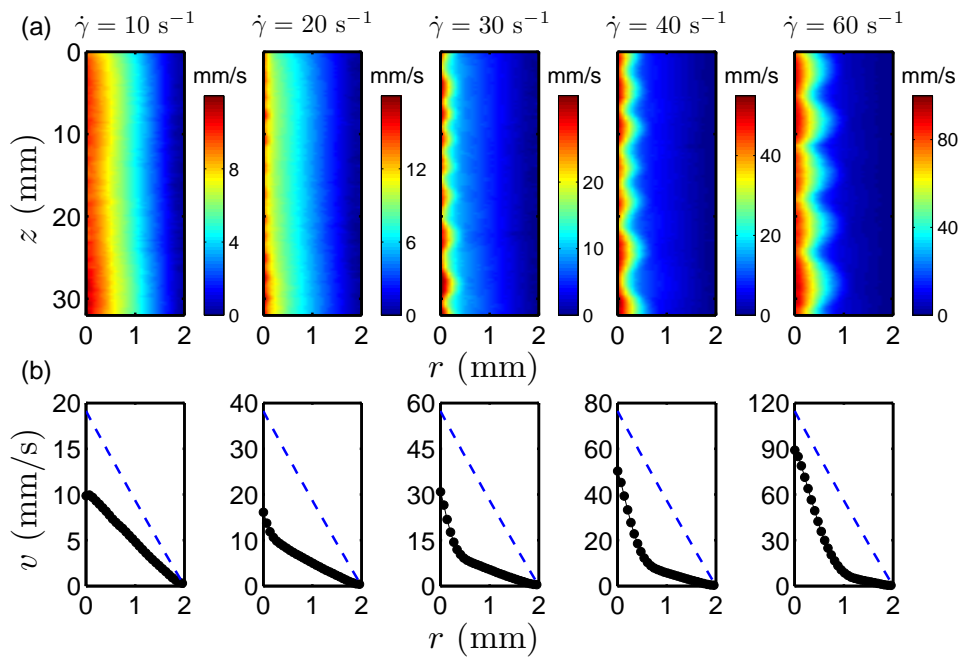}}
\caption{\label{f.steadystate}Steady state reached by a semidilute solution of wormlike micelles ([CTAB]=0.3 M, [NaNO$_3$]=0.4 M) seeded with hollow glass spheres at 1~wt.~\% at $T=30^\circ$C. (a)~Time-averaged velocity maps $\langle v(r,z,t)\rangle_t$ at different imposed shear rates $\gp$ indicated on the top row. $r$ corresponds to the radial distance to the inner rotating cylinder and $z$ denotes the vertical direction oriented downwards with $z=0$ being taken at about 6~mm from the top of the TC cell. Averages are taken over 8,000~consecutive pulses once steady state is reached (typically 120~s after flow inception). The time interval between two pulses depends on $\gp$ and is given by $1/(50\gp)$ so that the averaging time decreases from 16~s at 10~s$^{-1}$ to 2.7~s at 60~s$^{-1}$. (b)~Velocity profiles $\langle v(r,z,t)\rangle_{z,t}$ averaged over the whole height of the transducer array and over the same time interval as in (a). The blue dotted lines show the velocity profiles expected for a Newtonian fluid in the laminar regime.}
\end{center}
\label{f.steadystate}
\end{figure*}

\subsection{Ultrafast ultrasonic imaging of Taylor-Couette flows in wormlike micelles}

Building upon a previous one-dimensional ultrasonic velocimetry setup \cite{Manneville:2004a}, we have recently devised a new ultrasonic imaging system based on a multi-channel electronics that allows the simultaneous measurement of 128 velocity profiles over 32~mm along the vertical direction in the TC geometry \cite{Gallot:2013}. The reader is referred to Refs.~\cite{Manneville:2004a,Gallot:2013} for full technical details. In brief, this multi-channel setup is used to emit plane acoustic pulses by firing simultaneously 128 piezoelectric transducers arranged in a vertical array. These short pulses are backscattered either by the material microstructure itself or by acoustic contrast agents, here hollow glass spheres, seeding the material under investigation. The backscattered signals received on all 128 transducers are stored in large on-board memories then downloaded to a PC once the pulse sequence is completed. The recorded signals are further used to construct ultrasonic images of the scatterer distribution. By cross-correlating successive images corresponding to successive pulses and by averaging the correlations to increase signal-to-noise ratio, one recovers images of one component $v_y$ of the velocity field over the whole height and gap of the TC cell. This component corresponds to the projection of the velocity vector $\mathbf{v}=v_r\mathbf{e}_r+v_\theta\mathbf{e}_\theta+v_z\mathbf{e}_z=(v_r,v_\theta,v_z)$ [where ($\mathbf{e}_r$, $\mathbf{e}_\theta$, $\mathbf{e}_z$) are the unit vectors in cylindrical coordinates] along the acoustic propagation axis $y$ which is horizontal and makes an angle $\phi$ with the normal to the outer cylinder: $v_y=\cos\phi\, v_r + \sin\phi\, v_\theta$. The angle $\phi$ is inferred from an independent calibration step in a Newtonian fluid \cite{Gallot:2013}. Finally we define the measured velocity $v$ as
\begin{equation}
v=\frac{v_y}{\sin\phi}=v_\theta+\frac{v_r}{\tan\phi}\,,
\end{equation}
which coincides with the azimuthal velocity $v_\theta$ in the case of a purely orthoradial flow $\mathbf{v}=(0,v_\theta,0)$. In the more general case of a three-dimensional non-axisymmetric flow, it is important to keep in mind that our ultrasonic measurements of $v$ combine contributions from both azimuthal and radial velocity components.

Ultrasonic images are focused on the gap of a TC device with gap 2~mm, outer radius 25~mm, and height 60~mm. The curvature of such a cylindrical geometry leads to a stress inhomogeneity of 16~\% across the gap. In the present study both cylinders have smooth surfaces. By using a high ultrasonic center frequency of 15~MHz for the plane pulsed emissions and a sampling frequency of 160~MHz, we can access displacements of a few micro\-meters with a spatial resolution of about 100~$\mu$m. Thanks to the high temporal resolution of the ultrasonic scanner (up to 20,000 pulses per second), fast transients can be imaged as well as fluctuations in unstable flows.

In the following we focus on a well-studied semidilute wormlike micellar solution made of cetyltrimethylammonium bromide (CTAB) at a concentration of 0.3~M and sodium nitrate (NaNO$_3$) at 0.4~M in distilled water. Experiments were performed either at $T=28$ or 30$^\circ$C ($\pm 0.1^\circ$C), two highly studied temperatures for this surfactant system~\cite{Fardin:2012c,Lerouge:2006,Lerouge:2008}. The dynamics, referred to as ``shear-banding of category 2'' in Ref.~\cite{Fardin:2012d}, are very similar at those two temperatures. In order to get some acoustic contrast from the solution which is fully transparent to ultrasound, this standard system is seeded with 1~wt.~\% polydisperse hollow glass spheres (Potters Sphericel, mean diameter 6~$\mu$m, mean density 1.1~g.cm$^{-3}$) prior to surfactant dispersion. The addition of acoustic contrast agents does not significantly modify the rheological properties of the sample. Indeed, as can be checked in Fig.~\ref{f.rheology}(a), the linear viscoelastic moduli nicely follow the Maxwell model, which is typical of semidilute wormlike micelles \cite{Berret:2005,Cates:2006}, and the values of $G_0$ and $\tau$ are in agreement with the literature~\cite{Fardin:2012c,Lerouge:2006,Lerouge:2008}. The elastic modulus and relaxation time are respectively $G_0=238$~Pa and $\tau=0.18$~s at $T=30^\circ$C, and $G_0=239$~Pa and $\tau=0.25$~s at $T=28^\circ$C. Finally the flow curve displayed in Fig.~\ref{f.rheology}(b) for $T=30^\circ$C has a stress plateau at $\sigma^\star\simeq 160$~Pa that extends from $\gp_l\simeq 10$~\is\ to $\gp_h\simeq 120$~\is, which is also in line with previous works.

\subsection{Time-averaged flow in the stationary shear-banding regime}

\begin{figure}
\begin{center}
\resizebox{\columnwidth}{!}{\includegraphics{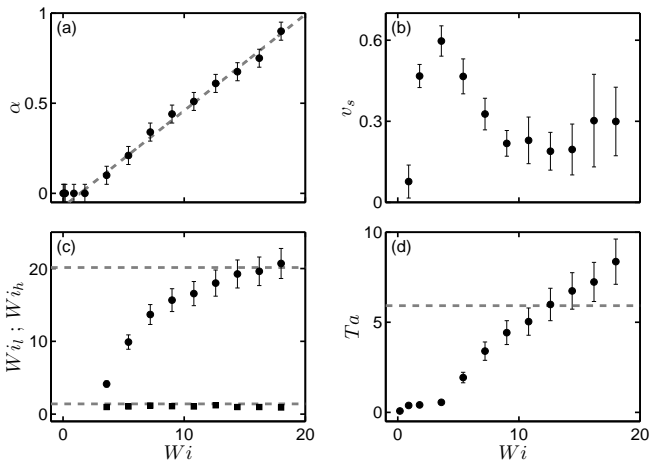}}
\caption{\label{f.analysis}Analysis of the space- and time-averaged velocity profiles in the shear-banding regime for $T=30^\circ$C. (a) Fraction $\alpha$ of the gap occupied by the high shear rate band plotted against the Weissenberg number $Wi=\tau\gp$. The proportion $\alpha$ is obtained via a piecewise linear fit of the velocity profiles averaged in space along $z$ and in time in the steady state. The dotted line is the best linear fit of the data in the shear-banding regime, $\alpha=(Wi-Wi_l)/(Wi_h-Wi_l)$, with $Wi_l\simeq 1.4$ and $Wi_h\simeq 20.1$. Error bars are $\pm 100~\mu$m. (b) Dimensionless averaged slip velocity at the rotor $v_s\equiv (v_0 - v(0))/v_0$, where $v_0$ is the imposed velocity of the rotor and $v(0)$ is the measured velocity of the fluid at the rotor. The average is along $z$ and time. The error bars are the standard deviations. (c) Local Weissenberg numbers in the low ($Wi_l\equiv\tau\gp_l$) and high ($Wi_h\equiv\tau\gp_h$) shear bands obtained from the piecewise fits of the averaged velocity profiles. Error bars are given by the experimental uncertainty related to the velocity measurements. The dashed lines are the values of $Wi_l$ and $Wi_h$ found in (a). (d) Local value of the viscoelastic Taylor number in the high shear rate band $Ta=\sqrt{\alpha(\gp) e/R_i} \tau\gp_h(\gp)$. The dashed line shows the threshold of the elastic instability predicted from the Upper Convected Maxwell model with hard boundaries~\cite{Larson:1990}.}
\end{center}
\end{figure}

\begin{figure}
\begin{center}
\resizebox{0.9\columnwidth}{!}{\includegraphics{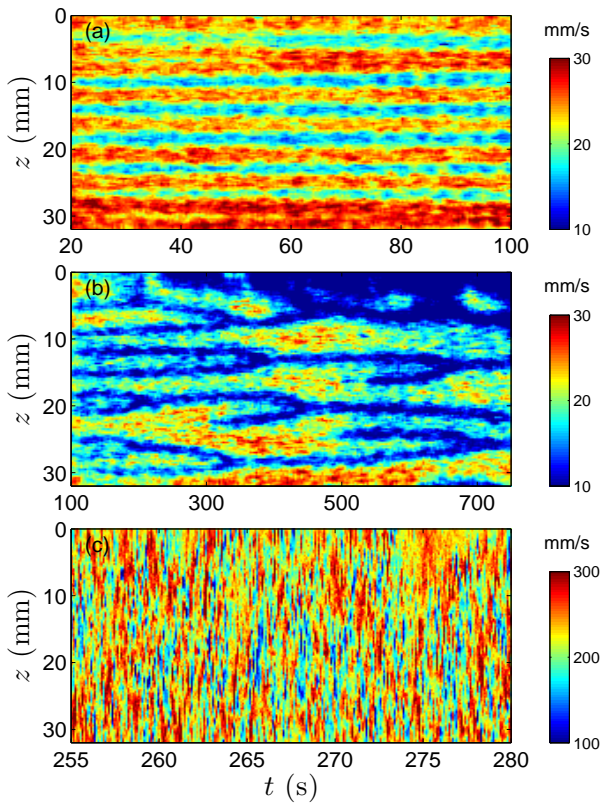}}
\caption{\label{f.spatiotp}Spatio-temporal diagrams of the velocity $v(r_0,z,t)$ for different shear rates. (a)~Quasi-stationary Taylor vortex flow at $\gp=30$~s$^{-1}$ for $r_0=0.14$~mm. (b)~Flame pattern at $\gp=75$~s$^{-1}$ for $r_0=1.15$~mm. (c)~Fully turbulent state at $\gp=140$~s$^{-1}$ for $r_0=0.14$~mm. Time $t=0$ corresponds to flow inception. Each data set is recorded once steady state is reached. Experiments performed at $T=28^\circ$C.}
\end{center}
\end{figure}

It is well-known that the CTAB-NaNO$_3$ system under study displays shear banding along the stress plateau \cite{Fardin:2012d,Lerouge:2008}. Therefore as a first check of previous results Fig.~\ref{f.steadystate}(a) shows the time-averaged velocity maps obtained under imposed shear rates ranging from 10~\is\ to 60~\is\ once steady state is achieved. We shall check in Fig.~\ref{f.spatiotp}(a) below that the flow remains stationary in this range of shear rates so that time-averaging is justified. When velocity maps are further averaged over the whole height of the transducer array, the resulting space- and time-averaged velocity profiles [see Fig.~\ref{f.steadystate}(b)] seem in good agreement with a standard shear-banding scenario although a large amount of wall slip is evident at the inner wall due to smooth boundaries \cite{Fardin:2012d,Lettinga:2009,Fardin:2012b}. More precisely the detailed analysis of these velocity profiles reported in Fig.~\ref{f.analysis} shows that the system exhibits a \textit{standard anomalous lever rule} \cite{Fardin:2012b}: (i)~the proportion $\alpha$ of high shear rate band within the gap of the TC cell increases linearly with the global shear rate or equivalently with the global Weissenberg number $Wi=\tau\gp$ [Fig.~\ref{f.analysis}(a)]; (ii)~the value of the local shear rate $\gp_l\simeq 6$~\is\ in the low shear rate band remains constant across the shear-banding regime and corresponds to the beginning of the plateau of the flow curve $\gp_l\simeq 10$~\is\ provided wall slip at the rotor is taken into account [Fig.~\ref{f.analysis}(c)]; (iii)~the shear rate in the high shear rate band is not constant but rather increases with the global shear rate until the value predicted by the lever rule is reached [Fig.~\ref{f.analysis}(c)]. This variation of the local high shear rate is correlated with wall slip at the rotor [Fig.~\ref{f.analysis}(b)]. This is fully consistent with previous one-dimensional measurements performed on the same system~\cite{Fardin:2012d}.

\begin{figure}
\begin{center}
\resizebox{0.8\columnwidth}{!}{\includegraphics{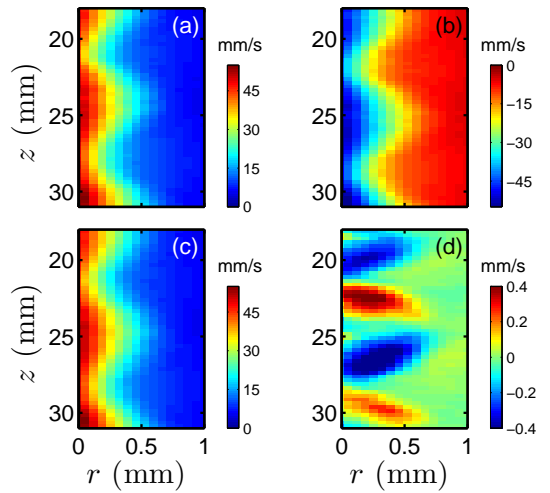}}
\caption{\label{f.rolls}Enlargements of time-averaged velocity maps recorded in the shear-banding regime at $T=30^\circ$C. (a)~$\langle v_+(r,z,t)\rangle_t$ and (b)~$\langle v_-(r,z,t)\rangle_t$ obtained by rotating the inner cylinder in opposite directions at $\gp=+40$~s$^{-1}$ and $-40$~s$^{-1}$ respectively. Averages are taken over 8,000~consecutive pulses separated by 0.5~ms once steady state is reached (typically 120~s after flow inception). (c) Azimuthal velocity map $v_\theta(r,z,t)$ and (d) radial velocity map $v_r(r,z,t)$ computed from (a) and (b) as explained in the text.}
\end{center}
\end{figure}

However thanks to two-dimensional imaging it is clearly revealed in Fig.~\ref{f.steadystate}(a) that velocity maps display vertical oscillations localized in the high shear rate band whatever the shear rate within the shear-banding regime. These oscillations are the signature of counter-rotating Taylor-like vortices localized in the aligned micelles due to an elastic instability as also shown by Lerouge and coworkers \cite{Fardin:2012d}. Here ultrasonic flow imaging allows us to go one step further and study in more details the structure of such vortices. Indeed by performing two different experiments at the same shear rate but with opposite rotation directions $+\gp$ and $-\gp$, one may combine the two corresponding velocity measurements $v_+$ and $v_-$ to recover the two horizontal flow components $v_\theta$ and $v_r$ using $v_\theta=(v_+-v_-)/2$ and $v_r=\tan\phi\,(v_++v_-)/2$. 

\begin{figure*}
\begin{center}
\resizebox{0.7\textwidth}{!}{\includegraphics{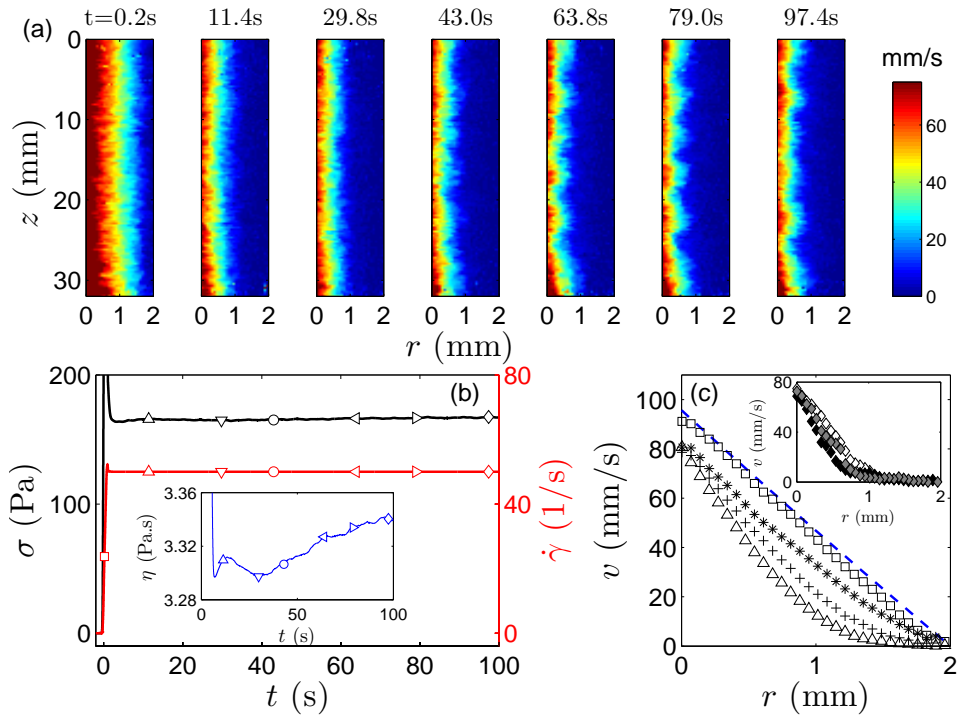}}
\caption{\label{f.startup}Start-up of shear at $\gp=50$~s$^{-1}$ in the shear-banding regime with Taylor-like vortices  at $T=28^\circ$C (see also supplementary movie~1). (a)~Velocity maps $v(r,z,t)$ at different times $t$ from flow inception indicated on the top row. Each map corresponds to an average over 50 pulses sent every 0.4~ms. (b)~Stress response $\sigma(t)$ (in black) recorded by the rheometer simultaneously to the velocity maps together with the instantaneous shear rate $\gp(t)$ (in red). The symbols indicate the times corresponding to the images shown in (a). Inset: apparent viscosity $\eta(t)=\sigma(t)/\gp(t)$. (c)~Velocity profiles $\langle v(r,z,t)\rangle_z$ averaged over the whole height of the transducer array and shown at $t=0.17$ ($\square$), 3.0 ($*$), 5.0 ($+$), and 9.0~s ($\triangle$). The blue dotted line shows the velocity profile expected for a Newtonian fluid in the laminar regime. Inset: velocity profiles $v(r,z,t)$ at $t=95.0$~s and for $z=22.4$ (white symbols, outflow boundary), 20.6 (gray symbols, in between outflow and inflow), and 18.9~mm (black symbols, inflow boundary).}
\end{center}
\end{figure*}

Obviously this procedure based on two independent experiments relies on the fact that the pattern of counter-rotating vortices is located at the exact same place in both experiments, which may not be the case due to sensitivity on initial conditions and on subtle details of the stress history. Moreover as seen in Fig.~\ref{f.steadystate}(a), the wavelength of the pattern is not perfectly uniform across the TC cell due to possible end effects. Yet Fig.~\ref{f.rolls} reports a case where the arrangements of vortices coincide in both $v_+$ and $v_-$ [see Figs.~\ref{f.rolls}(a) and (b)] so that $v_\theta$ and $v_r$ could be estimated without further data analysis [see Figs.~\ref{f.rolls}(c) and (d)]. These measurements confirm that for each wavelength of the undulations on the velocity map, there is a pair of counter-rotating vortices. Figure~\ref{f.rolls}(c) shows that the homogeneous, purely azimuthal base flow is strongly affected by the secondary flow. The order of magnitude of the perturbation to $v_\theta$ is about 15~mm.\is, much larger than the maximum radial velocity of about 0.5~mm.\is. Also revealed in Fig.~\ref{f.rolls}(d) is a striking asymmetry in the radial velocity map with the regions of positive $v_r$ (negative $v_r$ resp.) pointing downwards (upwards resp.) from the inner wall. We checked that this asymmetric structure is observed for all shear rates in the shear-banding regime and therefore cannot be attributed to a secondary instability far from the onset of primary elastic instability. To the best of our knowledge such a feature has not been reported in TC flows of Newtonian fluids nor polymers. For Newtonian fluids under pure inner rotation, the outward flow regions are a little more intense than the inward flow~\cite{Tagg:1994}. In polymer solutions undergoing purely elastic instability, the inward flow is much stronger than the outward flow as seen in the characteristic profile of a so-called ``diwhirl''~\cite{Groisman:1997}. But in both cases, inward or outward flows do not show any asymmetry such as observed here. Whether this feature is inherent to the presence of the soft interface between shear bands or whether it is characteristic of elastically unstable flows in micellar fluids remains an open question which should be addressed both theoretically and experimentally. Note that the oscillations of the interface positions when it is closest to the inner wall~\cite{Lerouge:2008} and the peculiar rope-like structure of spatiotemporal diagrams of flow tracers~\cite{Fardin:2009} reported by Lerouge \textit{et al.} may be linked to this peculiar structure of the secondary flow field.  

Finally the presence of the soft boundary between the unstable high shear rate band and the stable low shear rate band has been shown to influence the spatiotemporal dynamics of the vortex flow~\cite{Fardin:2012c}. To the first order of approximation, it has an impact through the modification of the instability threshold~\cite{Fardin:2011}. In Fig.~\ref{f.analysis}(d) we have computed the elastic Taylor number in the unstable band $Ta=\sqrt{e \alpha/R_i} Wi_h$, where $Wi_h\equiv\tau\gp_h(\gp)$~\cite{Fardin:2011}. Clearly the theoretical threshold prediction for the Upper-Convected Maxwell model with hard boundaries~\cite{Larson:1990} does not fit the data, since shear-banded flows below the threshold have been shown to be unstable. On a finer level the boundary could also generate interfacial instability modes~\cite{Fielding:2010,Nicolas:2012}. 
 
\subsection{Transient measurements during start-up in the shear-banding regime}

The time-resolved capabilities of our ultrasonic technique are illustrated in Fig.~\ref{f.startup} through measurements during flow inception at a shear rate well within the stress plateau (see also supplementary movie~1). At short times ($t\lesssim 10$~s), the flow remains homogeneous over the whole height of the TC device. Velocity profiles averaged over the vertical direction $z$ show the progressive radial separation of the flow into two shear bands with the highly sheared aligned state at the vicinity of the inner rotating cylinder [see Fig.~\ref{f.startup}(c)]. These results are consistent with previous findings on other wormlike micellar systems \cite{Lerouge:2010,Lettinga:2009,Hu:2005}.

\begin{figure}
\begin{center}
\resizebox{0.9\columnwidth}{!}{\includegraphics{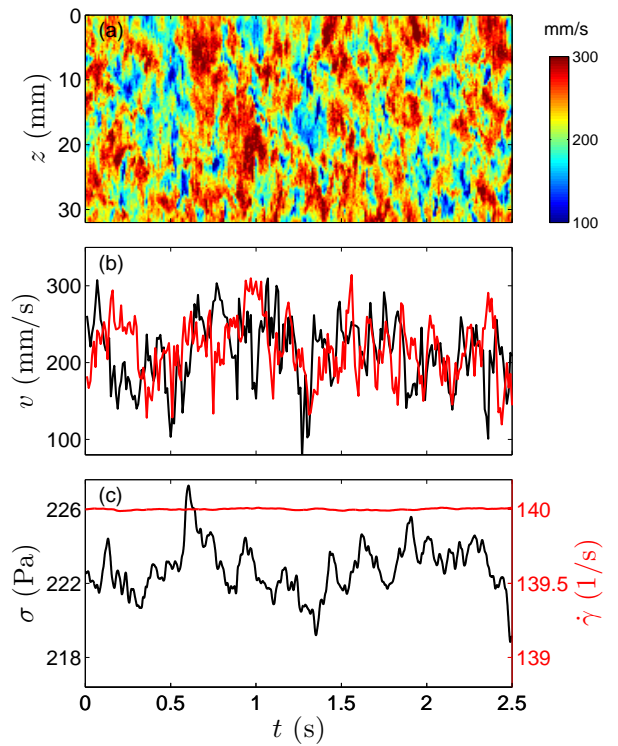}}
\caption{\label{f.turbul}Elastic turbulence at $\gp=140$~s$^{-1}$ over a short time window of duration 2.5~s starting 63~s after flow inception  (see also supplementary movie~2). (a)~Spatio-temporal diagram of the velocity $v(r_0,z,t)$ for $r_0=0.14$~mm. (b)~Velocity time series $v(r_0,z_0,t)$ for $r_0=0.14$~mm and $z_0=9.9$~mm (black line) and $z_0=19.9$~mm (red line). (c)~Rheological data, shear rate $\gp(t)$ (red line) and shear stress $\sigma(t)$ (black line), recorded by the rheometer simultaneously to the velocity. Velocity measurements correspond to averages over 7~ms (50 pulses separated by 140~$\mu$s). Experiment performed at $T=28^\circ$C.}
\end{center}
\end{figure}

Moreover the fast imaging technique used in the present work allows us to visualize the growth of a subsequent instability that develops in the highly sheared band (for $r\lesssim 1$~mm). Indeed for $t\gtrsim 30$~s, the vertical symmetry is lost and the interface between shear bands oscillates along the $z$ direction while the velocity profile in the low shear band remains unaffected [see inset of Fig.~\ref{f.startup}(c)]. This corresponds to the onset and growth of the elastic instability within the high shear rate band that eventually results in the steady-state described in the previous paragraph. Note also that the onset of Taylor-like vortices induces an additional dissipation that shows as an increase of the viscosity in the inset of Fig.~\ref{f.startup}(b). This increased dissipation was reported both for the inertial TC instability in Newtonian fluids~\cite{Taylor:1936} and for the elastic TC instability in polymers~\cite{Larson:1990} and in wormlike micelles \cite{Fardin:2012d,Lerouge:2008}.

\subsection{Complex dynamics in steady-state at large shear rates}

For about a decade semidilute systems have been known to present complex temporal fluctuations along the stress plateau and above \cite{Becu:2004,Becu:2007,Lopez:2004}. Thanks to rheo-optical measurements this has been attributed to secondary instabilities that develop from the primary (stationary) elastic instability \cite{Fardin:2009,Fardin:2012d}. When the shear rate is increased the stable Taylor-like vortices become unstable and show complex dynamical regimes reminiscent of phase turbulence \cite{Lerouge:2006,Lerouge:2008}. Eventually Taylor-like vortices give way to fully-developed elastic turbulence \cite{Fardin:2010}. Depending on the system under study and on the shearing geometry this sequence of events may be observed in the shear-banding regime~\cite{Fardin:2012c,Fardin:2012b}.

While previous works were based on optical visualizations the present ultrasonic imaging technique allows for a quantitative analysis of fluctuating velocity maps. In Fig.~\ref{f.spatiotp}(a) we start by checking that Taylor-like vortices remain stable in the low-shear regime in the stress plateau. Except for a fusion event at $t\simeq 60$~s where two vortices located at $z\simeq 5$--10~mm merge, the velocity close to the rotor remains broadly time-independent once the vortices have grown has described above in Fig.~\ref{f.startup}. For larger shear rates, e.g. $\gp=75$~s$^{-1}$ as shown in Fig.~\ref{f.spatiotp}(b), annihilation and creation of vortices are observed, which leads to a typical ``flame pattern'' \cite{Muller:2008,Fardin:2012c}. At $\gp=140$~s$^{-1}$ elastic turbulence shows up as fast and heterogeneous fluctuations [see Fig.~\ref{f.spatiotp}(c)]. 

Figure~\ref{f.turbul} provides a deeper view on the elastic turbulence at $\gp=140$~s$^{-1}$ through velocity measurements with a high temporal resolution over 2.5~s  (see also supplementary movie~2). The spatiotemporal diagram of Fig.~\ref{f.turbul}(a) plotted for a given radial position $r_0$ close to the rotor reveals significant spatial coherence of the flow field in the form of large patches of high velocities that generally propagate downwards in the TC cell. This shows as noticeable correlation between the two signals $v(r_0,z_0,t)$ shown in Fig.~\ref{f.turbul}(b) for two vertical positions separated by 10~mm. The fact that these time series seem decorrelated over some time intervals (e.g. $t\simeq 0.2$ or 1~s) suggests that these turbulent patches also move in the radial direction. Finally Fig.~\ref{f.turbul}(c) shows that the shear rate $\gp$ is held fixed up to better than 0.05\% in spite of fluctuations of the stress of the order of 2\%. As expected for a global measurement most fluctuations are smoothed out in the stress signal. Still a stress minimum (e.g. at $t\simeq 0.4$~s) corresponds to smaller velocities while intense turbulent patches appear to be related to stress excursions towards larger values (e.g. at $t\simeq 0.6$ or 0.9~s). A more detailed analysis of spatio-temporal fluctuations in the turbulent regime is left for future work.

\section{Conclusions and perspectives}

In this paper we have shown that surfactant wormlike micelles appear as a model system to study instabilities induced by shear in complex fluids, even at vanishingly small Reynolds numbers. Ultrasonic imaging in a standard semidilute wormlike micellar system fully confirms previous works based on rheo-optics. We have provided space- and time-resolved velocity data based on a new multi-channel ultrasonic scanner coupled with a standard rheometer. When averaging velocity maps in space and time a classical shear-banding scenario is observed over a large part of the stress plateau. However spatially resolved measurements reveal the presence of a secondary flow in the form of counter-rotating vortices similar to the Taylor vortices found in the inertial Taylor-Couette instability yet with a significantly different internal structure. This vortical flow remains stationary at low shear rates in the stress plateau and gives way to complex fluctuating vortices and eventually to a turbulent state when the shear rate is increased.

The interplay between shear banding and elastic instabilities in semidilute wormlike micelles gives rise to original phenomena. Future experiments will focus on trying to discriminate between elastic and inertial vortices, on providing a more detailed view of the internal structure of those vortices, and on performing a statistical analysis of elastic turbulence. We have also emphasized that dilute and concentrated regimes respectively display shear-induced structures and yield stress behaviour. Whether such an interplay with elastic instabilities as in semidilute systems is present in those other wormlike micellar systems remains an open issue that will be addressed in upcoming experiments. Of particular interest are (i)~the transition to dilute systems where elastic or inertio-elastic instabilities are expected without shear banding and (ii)~the transition to yield stress fluids where the peculiar relaxation dynamics of wormlike micelles may allow us to study the limit where both characteristic shear rates $\gp_l$ and $\gp_h$ in Fig.~\ref{f.rheol} successively go to zero.
 
\begin{acknowledgement}
The authors thank J.F.~Berret, T.~Divoux, T.~Gallot, S.~Lerouge, and N.~Ta\-ber\-let for fruitful discussions. This work was funded by the European Research Council under the European Union's Seventh Framework Programme (FP7/2007-2013) / ERC grant agreement n$^\circ$~258803. SM is also affiliated with and funded by Institut Universitaire de France.

\end{acknowledgement}

\end{document}